\documentclass[10pt]{iopart}

\usepackage[english]{babel}
\usepackage{graphicx}
\usepackage{dcolumn}
\usepackage{bm}
\usepackage{iopams}
\def\pp{p_{\rm Ch}}
\def\rr{\rho_{\rm Ch}}
\def\ww{w_{\rm Ch}}
\begin{document}

\newcommand{\h}{\mathcal{H}}

\title[Gauge--invariant analysis of perturbations in Chaplygin gas
unified models]{Gauge--invariant analysis of perturbations in
Chaplygin gas unified models of dark matter and dark energy}

\author{V Gorini$^{1,2}$, A Y Kamenshchik$^{3,4}$, U
Moschella$^{1,2}$, O F Piattella$^1$ and A A Starobinsky$^{4,5}$}

\address{$^1$ Dipartimento di Scienze Fisiche e Matematiche,
Universit\`a dell'Insubria,
Via Valleggio 11, 22100 Como, Italy}
\address{$^2$ INFN, sez. di Milano, Via Celoria 16, 20133 Milano, Italy}
\address{$^3$ Dipartimento di Fisica and INFN,
Via Irnerio 46, 40126 Bologna, Italy}
\address{$^4$ L.D. Landau Institute for Theoretical Physics,
Russian Academy of Sciences, Kosygin str. 2,
119334 Moscow, Russia}
\address{$^5$ Yukawa Institute for Theoretical Physics, Kyoto University,
Kyoto 606-8502, Japan}

\eads{\mailto{vittorio.gorini@uninsubria.it},
\mailto{kamenshchik@bo.infn.it},
\mailto{ugo.moschella@uninsubria.it},
\mailto{oliver.piattella@uninsubria.it} and
\mailto{alstar@landau.ac.ru}}

\begin{abstract}
We exploit the gauge--invariant formalism to analyse the
perturbative behaviour of two cosmological models based on the
generalized Chaplygin gas describing both dark matter and dark
energy in the present Universe. In the first model we consider the
generalized Chaplygin gas alone, while in the second one we add a
baryon component to it. We extend our analysis also into the
parameter range $\alpha > 1$, where the generalized Chaplygin gas
sound velocity can be larger than that of light.

In the first model we find that the matter power spectrum is
compatible with the observed one only for $\alpha < 10^{-5}$, which
makes the generalized Chaplygin gas practically indistinguishable
from $\Lambda$CDM.

In the second model we study the evolution of inhomogeneities of the
baryon component. The theoretical power spectrum is in good
agreement with the observed one for almost all values of $\alpha$.
However, the growth of inhomogeneities seems to be particularly
favoured either for sufficiently small values of $\alpha$ or for
$\alpha \gtrsim 3$. Thus, it appears that the viability of the
generalized Chaplygin gas as a cosmological model is stronger when
its sound velocity is superluminal. We show that in this case the
generalized Chaplygin gas equation of state can be changed in an
unobservable region in such a way that its equivalent $k$-essence
microscopical model has no problems with causality.

\end{abstract}

\noindent{\it Keywords\/}: Chaplygin gas, Cosmological perturbation
theory, gauge--invariant formalism, large scale structure power
spectrum

\submitto{{\it JCAP}}

\maketitle

\section{Introduction}\label{Intro}

Together with quintessence (a scalar field with some potential
minimally coupled to gravity) \cite{PR88,W88,Cald,S98} and other
physical and geometrical (modified gravity) models of dark energy
(see, e.g., the reviews \cite{SS00,SS06} for exact definitions), the
generalized Chaplygin gas \cite{Kam2} (hereafter gCg) is one widely
studied model among those proposed to describe the observed
accelerated expansion of the universe \cite{Riess,Perlmutter}. In
its generalized form, the Chaplygin gas is a barotropic fluid with
the following equation of state:
\begin{equation}\label{gcgeos}
\pp  = - A\rr^{-\alpha},
\end{equation}
where $A$ and $\alpha$ are positive constants (for the original
Chaplygin gas $\alpha = 1$). The generalized equation of state
(\ref{gcgeos}) has been introduced in \cite{Kam2} and analysed in
\cite{Bilic,bentobertosen,Gor} and in many other subsequent papers.
In contrast to many models describing dark energy alone, the gCg
gives a unified description of dark matter and dark energy,
enrolling itself in the class of so-called quartessence or unified
dark matter (UDM) cosmological models. It allows to interpolate
between a dust--dominated phase of the evolution of the Universe in
the past and an accelerated one at recent time, see Eq.
(\ref{gcgrhoevo}) below. That is why the gCg has attracted much
attention in cosmology.

The evolution of the energy density as a function of the cosmic
scale factor $a(t)$ of the
Friedmann--Lema\^{i}tre--Robertson--Walker (FLRW) cosmological model
is easily obtained from (\ref{gcgeos}) and from the energy
conservation equation:
\begin{equation}\label{gcgrhoevo}
\rr = \left(A + \frac{B}{a^{3(1 + \alpha)}}\right)^{\frac{1}{1 +
\alpha}},
\end{equation}
where $B$ is an integration constant usually chosen to be positive
(if $B < 0$, the weak energy condition is violated and phantom
cosmology takes place, see for example \cite{Zhang,Bou}).

 From (\ref{gcgeos}) and (\ref{gcgrhoevo}), the parameter $\ww \equiv
p_{\rm Ch}/\rho_{\rm Ch}$ and the square sound velocity $c_{\rm
sCh}^{2}$ have the following expressions: \numparts
\begin{eqnarray}
w_{\rm Ch} &=& -\left[1 + \frac{B}{A}\frac{1}{a^{3\left(\alpha +
1\right)}}
\right]^{-1}, \label{gcgw}\\
c_{\rm sCh}^{2} &\equiv& {\frac {{\rm d} \pp}{{\rm d} \rr}} =
-\alpha w_{\rm Ch}. \label{gcgcs}
\end{eqnarray}
\endnumparts
Note that for $\alpha > 1$, $c_{\rm sCh}^{2}$ may exceed the
velocity of light (set to unity in our notations) in the course of
the recent or future evolution of the Universe, while it was
non-relativistic at large redshifts, during the matter--dominated
stage, for all values of $\alpha$.

Cosmological models based either on one or on more than one fluid
are essentially different. The reason for this is that the various
fluid components interact indirectly via geometry and may also
interact by direct energy exchange. Direct non-gravitational
interactions are typically neglected but the indirect ones are
always present and can give rise to totally different evolutions,
especially at the perturbative level. We will see how this works for
the gCg together with baryons.

The pure gCg, namely the cosmological model based on the gCg alone,
has passed many tests of standard cosmology. Quite recently, updated
constraints for the gCg parameters have been published
\cite{Puxun,WY07,Davis}. However, the behaviour of the gCg under
perturbations is still problematic \cite{Bean, Sandvik}. In
\cite{Sandvik} the study of the power spectrum of large scale
structures seems to indicate that the best fit value of $\alpha$ is
very close to zero, rendering the gCg indistinguishable from
$\Lambda$CDM [the latter is indeed the limit of gCg for $\alpha \to
0$, as one can infer from (\ref{gcgH}) below]. This feature has to
be attributed to the gCg sound velocity which, during the cosmic
evolution, grows from 0 to $\sqrt{\alpha}$ driving inhomogeneities
to oscillate (if $\alpha > 0$) or to blow--up (if $\alpha < 0$).
This characteristic seems to be common to all UDM models
\cite{Sandvik} which, as a consequence, might appear to be ruled
out.

However, as shown in \cite{Beca}, it appears that adding a baryon
component to the gCg the problem disappears and the model is in
agreement with observations. But the debate still goes on and the
authors of \cite{Sandvik} have corroborated their claim about ruling
out gCg and, in general, all quartessence models through
gravitational lensing measurements on the basis of the current value
of the cosmological parameter $\sigma_{\rm 8}$ (the rms mass
dispersion on a sphere of radius $8h^{-1}$ $Mpc$).

Still, the parameter $\sigma_{\rm 8}$ need not necessarily be a
confident discriminant among different cosmological models since it
is clearly a quantity which is strongly affected by non--linear
effects, while all the calculations performed in \cite{Sandvik} are
carried out in the linear regime. This point was also raised much
earlier in \cite{peacock} in a more general context.

In the present paper we investigate the issue of cosmological
perturbations in gCg models using the gauge--invariant formalism
\cite{Bardeen}. We consider both the gCg alone and in the presence
of baryons, confirming the results of \cite{Sandvik} and
\cite{Beca}. In addition, the gCg seems to favour structure
formation when its sound velocity is superluminal, in particular
when $\alpha \gtrsim 3$.

The structure of the paper is the following:
\begin{enumerate}
\item In section \ref{Giform} we briefly outline Bardeen gauge--invariant
formalism and derive the equations that we will numerically solve in
the rest of the paper.
  \item In section \ref{gcgruled} we investigate structure formation
  in a gCg--dominated universe and display the role of the sound velocity.
\item In section \ref{gcgpowers} we calculate the theoretical power spectrum
  in a gCg--dominated universe and compare it with the observed one.
  \item In section \ref{gcgandbar} we consider a two--fluid model based on
  gCg in presence of baryons. For this variant of the model we perform the
  same analysis of the previous two sections and compare the
  results.
\item In section \ref{concl} we present our conclusions and discuss in more
detail why a possible superluminal sound velocity of the gCg is not
prohibited by causality arguments.
\item For completeness, we add an appendix in which, using the technique developed in \cite{SolovStar} we
write down the exact solution for perturbations of dust-like matter
and generalized Chaplygin gas in the Newtonian approximation at the
matter--dominated stage.
\end{enumerate}

\section{The Gauge--Invariant Formalism}\label{Giform}

The cosmological perturbations issue was first tackled and studied
by Lifshitz in 1946 in the synchronous gauge \cite{Lifshitz} (see
also \cite{Lifkal}). In 1980 Bardeen developed gauge invariant
perturbation theory \cite{Bardeen} by constructing suitable
combinations of metric and stress--energy tensor perturbations which
are invariant under a generic gauge transformation. The
gauge--invariant formalism has been used and reviewed by many
authors. Here we follow the notation of \cite{Mukhanov} and study
only scalar perturbations. The first--order Einstein equations have
the following form:
\begin{equation}\label{Muksys}
\fl \left\{
\begin{array}{l}
\Delta\Psi - 3\mathcal{H}\left(\mathcal{H}\Phi + \Psi'\right) +
3\mathcal{K}\Psi = a^{2}\delta\rho\\
\mathcal{H}\Phi + \Psi' = a\left(\rho +
p\right)V\\
\left[\Psi'' + \mathcal{H}\Phi' + \left(2\mathcal{H}' +
\mathcal{H}^{2}\right)\Phi + 2\mathcal{H}\Psi' - \mathcal{K}\Psi +
\frac{1}{2}\Delta\left(\Phi - \Psi\right)\right]\delta^{i}_{j} - \\
- \frac{1}{2}\gamma^{ik}\left(\Phi - \Psi\right)_{|kj} = a^{2}\delta
p\delta^{i}_{j} - \sigma^{|i}_{|j},
\end{array}
\right.
\end{equation}
where $a(\eta)$ is the scale factor as a function of the conformal
time $\eta$. Its present value is normalized to unity;
$\mathcal{H}(\eta) = a'/a$, where the prime denotes derivation with
respect to the conformal time; $\Phi(\bi{x},\eta)$ and
$\Psi(\bi{x},\eta)$ are the  Bardeen gauge--invariant potentials;
$\delta\rho(\bi{x},\eta)$ and $\delta p(\bi{x},\eta)$ are the
gauge--invariant expressions of the perturbations of the energy
density and pressure; $V(\bi{x},\eta)$ is the gauge--invariant
expression of the scalar potential of the velocity field;
$\sigma(\bi{x},\eta)$ represents the shear. Finally, $\gamma_{ij}$
($i,j = 1,2,3$) is the spatial part of the FLRW metric and
$\mathcal{K} = 0, \pm 1$ is its curvature parameter which
corresponds respectively to flat, close and open geometry. The
vertical bar denotes covariant derivation with respect to
$\gamma_{ij}$. Units are chosen such that $4\pi G = c = 1$. See
\cite{Mukhanov} for a detailed derivation of system (\ref{Muksys}).

We introduce two assumptions which simplify (\ref{Muksys}):
\begin{itemize}
  \item We assume $\mathcal{K} = 0$;
  \item We neglect shear perturbations, namely we assume $\sigma = 0$.
This implies $\Phi = \Psi$.
\end{itemize}
For a cosmological model based on $N$ non--interacting fluids we
write for the background quantities:
\begin{equation}\label{rp}
\begin{array}{lr}
\rho = \sum_{i = 1}^{N}\rho_{i}, & p = \sum_{i = 1}^{N}p_{i},
\end{array}
\end{equation}
while for the perturbations in the linear regime:
\begin{equation}\label{deltarp}
\begin{array}{ll}
\delta\rho = \sum_{i = 1}^{N}\delta\rho_{i}, & \delta p = \sum_{i =
1}^{N}\delta p_{i}.
\end{array}
\end{equation}
The mixed time--space component of the perturbed energy--momentum
tensor will look like $\sum_{i = 1}^{N}\left(\rho_{i} +
p_{i}\right)V_{i}$, since each fluid will contribute with its own
velocity.

Adiabatic perturbations of a barotropic fluid are characterized by
the following equation of state:
\begin{equation}\label{adpert}
\delta p_{i} = c_{{\rm s}i}^{2} \delta\rho_{i},
\end{equation}
where
\begin{equation}\label{cssq}
c_{{\rm s}i}^{2} \equiv \frac{{\rm \partial}p_{i}}{{\rm
\partial}\rho_{i}}
\end{equation}
is the sound velocity at constant entropy for the generic component
$i$. In the cosmological perturbation theory it is customary to deal
with the density contrast:
\begin{equation}\label{denscont}
\delta_{i} \equiv \frac{\delta\rho_{i}}{\rho_{i}},
\end{equation}
where $\rho_{i}$ is the background energy density of the component
$i$. Since we work in the linear regime it is useful to take the
spatial Fourier transform of (\ref{Muksys}) which allows us to treat
each mode independently. With the above assumptions and trading the
conformal time for the scale factor $a$, system (\ref{Muksys}) for a
generic multi--fluid model is rewritten as follows:
\begin{equation}\label{Muksys2}
\fl \left\{
\begin{array}{l}
-k^{2}\Phi - 3a\mathcal{H}^{2}\dot{\Phi} - 3\mathcal{H}^{2}\Phi = a^{2}
\sum_{i=1}^{N}\rho_{i}\delta_{i}\\
\mathcal{H}\Phi + a\h\dot{\Phi} = a\sum_{i =
1}^{N}\left(\rho_{i} + p_{i}\right)V_{i}\\
\left(a\mathcal{H}\right)^{2}\ddot{\Phi} + \left(4a\mathcal{H}^{2} +
a^{2}\h\dot{\h}\right)\dot{\Phi} + \left(2a\h\dot{\h} +
\mathcal{H}^{2}\right)\Phi = a^{2}\sum_{i=1}^{N}c_{{\rm
s}i}^{2}\rho_{i}\delta_{i},
\end{array}
\right.
\end{equation}
where the dot denotes derivation with respect to the scale factor
$a$ (for the sake of simplicity we adopt the same notation for the
Fourier transformed quantities as for the original ones).

If $N = 1$, the system is determined and can be solved by
eliminating $\delta$ and by extracting the following second order
equation for $\Phi$:
\begin{equation}\label{eqPhi}
\ddot{\Phi} + \left(\frac{\dot{\h}}{\h} + \frac{4}{a} +
\frac{3c_{\rm s}^{2}}{a}\right)\dot{\Phi} +
\left(2\frac{\dot{\h}}{a\h} + \frac{1 + 3c_{\rm s}^{2}}{a^{2}} +
\frac{k^{2}c_{\rm s}^{2}}{a^{2}\h^{2}}\right)\Phi = 0.
\end{equation}
Once solved, we can use the solution of (\ref{eqPhi}) in the first
equation of (\ref{Muksys2}) to find the solution for $\delta$.

Instead, if $N > 1$ we need $2(N - 1)$ additional equations to make
(\ref{Muksys2}) determined. These equations should take into account
the possible interactions and the energy exchanges among the
different fluids. Since we have assumed that they are not mutually
interacting directly, the required additional equations to be added
can be simply chosen as the energy conservation equations, $\delta
T^{\mu}_{\nu ;\mu} = 0$, given separately for each further
component. Namely the continuity equation $\delta T^{\mu}_{0 ;\mu} =
0$:
\begin{equation}\label{conteq}
\dot{\delta}_{i} + \frac{3}{a}\left(c_{{\rm s}i}^{2} -
w_{i}\right)\delta_{i} - 3\dot{\Phi}\left(1 + w_{i}\right) +
\frac{k^{2}}{a^{2}\h}\left(1 + w_{i}\right)V_{i} = 0
\end{equation}
and the Euler equation $\delta T^{\mu}_{l;\mu} = 0$:
\begin{equation}\label{Euleq}
\left[\left(\rho_{i} + p_{i}\right)V_{i}\right]\dot{} +
\frac{3}{a}\left(\rho_{i} + p_{i}\right)V_{i} - \frac{c_{{\rm
s}i}^{2}\rho_{i}}{\h}\delta_{i} - \frac{\left(\rho_{i} +
p_{i}\right)}{\h}\Phi = 0,
\end{equation}
where $w_{i} \equiv p_{i}/\rho_{i}$. Both (\ref{conteq}) and
(\ref{Euleq}) are Fourier transformed, that is why there is no $l$
subscript in (\ref{Euleq}).

It is possible to write the combination of (\ref{Muksys2}),
(\ref{conteq}) and (\ref{Euleq}) as a system of $N$ second order
equations, one for each $\delta_{i}$. One needs to derive
(\ref{conteq}), then to express $\ddot{\Phi}$, $\dot{\Phi}$, $V_{i}$
and $\dot{V}_{i}$ in terms of the other $\delta_{i}$'s through
(\ref{Muksys2}), (\ref{Euleq}) and again (\ref{conteq}). In the case
$N = 2$ the resulting system has the following form:
\begin{equation}\label{gensys}
\left\{
\begin{array}{l}
\ddot{\delta}_{\rm 1} + \left(\frac{\dot{\h}}{\h} + \frac{A_{\rm
1}}{\h}\right)\dot{\delta}_{\rm 1} + \frac{B_{\rm
1}}{\h}\dot{\delta}_{\rm 2} + \frac{C_{\rm 1}}{\h^{2}}\delta_{\rm 1}
+ \frac{D_{\rm 1}}{\h^{2}}\delta_{\rm 2} = 0\\ \\
\ddot{\delta}_{\rm 2} + \left(\frac{\dot{\h}}{\h} + \frac{A_{\rm
2}}{\h}\right)\dot{\delta}_{\rm 2} + \frac{B_{\rm
2}}{\h}\dot{\delta}_{\rm 1} + \frac{C_{\rm 2}}{\h^{2}}\delta_{\rm 2}
+ \frac{D_{\rm 2}}{\h^{2}}\delta_{\rm 1} = 0,
\end{array}
\right.
\end{equation}
where the coefficients have the following expressions:
\begin{eqnarray}\label{coeff}
\fl A_{\rm 1} = 2\frac{\h}{a} + 3\frac{\h}{a}\left(c_{\rm s1}^{2} -
2w_{\rm 1}\right) \nonumber\\
\fl - 3a\h\left(\rho_{\rm 1} + p_{\rm
1}\right)\frac{3\h^{2}\left(3c_{\rm s1}^{2} - 1\right) +
k^{2}\left(3c_{\rm s1}^{2} + 1\right) + 6\left(\h^{2} -
\h\dot{\h}a\right)}{k^{4} + \left(3\h^{2} +
k^{2}\right)\left(\h\dot{\h}a - \h^{2}\right)},\\
\fl B_{\rm 1} = - 3a\h\rho_{\rm 2}\left(1 + w_{\rm
1}\right)\frac{3\h^{2}\left(3c_{\rm s1}^{2} - 1\right) +
k^{2}\left(3c_{\rm s1}^{2} + 1\right) + 6\left(\h^{2} -
\h\dot{\h}a\right)}{k^{4} + \left(3\h^{2}
+ k^{2}\right)\left(\h\dot{\h}a - \h^{2}\right)},\\
\fl C_{\rm 1} = \frac{k^{2}c_{\rm s1}^{2}}{a^{2}} +
\frac{3}{a^{2}}\left(\dot{\h}\h a + 4\h^{2}\right)\left(c_{\rm
s1}^{2} - w_{\rm 1}\right) + 3\frac{\h^{2}}{a}\left(c_{\rm
s1}^{2}\right)\dot{} - 3\left(1 +
w_{1}\right)c_{\rm s1}^{2}\rho_{\rm 1} + \nonumber \\
\fl 3\h^{2}\left(2 + 3c_{\rm s1}^{2}\right)\left(\rho_{\rm 1} +
p_{\rm 1}\right)\frac{k^{2} - 3\left(c_{\rm s1}^{2} - w_{\rm
1}\right)\left(3\h^{2} + k^{2}\right)}{k^{4} + 3\left(3\h^{2}
+ k^{2}\right)\left(\h\dot{\h}a - \h^{2}\right)} \nonumber\\
\fl -\left(k^{2} + 3\h^{2} - 6\h\dot{\h}a\right)\left(\rho_{\rm 1} +
p_{\rm 1}\right)\frac{k^{2} + 3\left[\h\dot{\h}a - \h^{2} -
3\h^{2}\left(c_{\rm s1}^{2} - w_{\rm 1}\right)\right]}{k^{4}
+ 3\left(3\h^{2} + k^{2}\right)\left(\h\dot{\h}a - \h^{2}\right)},\\
\fl D_{\rm 1} = -3\left(1 + w_{\rm 1}\right)c_{\rm s2}^{2}\rho_{\rm
2}+ \nonumber \\
\fl 3\h^{2}\left(2 + 3c_{\rm s1}^{2}\right)\rho_{\rm 2}\left(1 +
w_{\rm 1}\right)\frac{k^{2} - 3\left(c_{\rm s2}^{2} - w_{\rm
2}\right)\left(3\h^{2} + k^{2}\right)}{k^{4} + 3\left(3\h^{2}
+ k^{2}\right)\left(\h\dot{\h}a - \h^{2}\right)} \nonumber\\
\fl -\left(k^{2} + 3\h^{2} - 6\h\dot{\h}a\right)\rho_{\rm 2}\left(1
+ w_{\rm 1}\right)\frac{k^{2} + 3\left[\h\dot{\h}a - \h^{2} -
3\h^{2}\left(c_{\rm s2}^{2} - w_{\rm 2}\right)\right]}{k^{4} +
3\left(3\h^{2} + k^{2}\right)\left(\h\dot{\h}a - \h^{2}\right)}.
\end{eqnarray}
The coefficients $A_{\rm 2}$, $B_{\rm 2}$, $C_{\rm 2}$ and $D_{\rm
2}$ have the same form, but with the interchange $1 \leftrightarrow
2$ in the subscripts.

As shown in \cite{SolovStar}, at the matter--dominated regime, i.e.
when $p_{\rm 1} \ll \rho_{\rm 1}$, $p_{\rm 2} \ll \rho_{\rm 2}$ and
$c_{\rm s1}, c_{\rm s2} \ll 1$, the above coefficients are far
simpler and system (\ref{gensys}) can be rewritten in the following
quasi--Newtonian form:
\begin{equation}\label{gensyssimplified}
\left\{
\begin{array}{l}
\ddot{\delta}_{\rm 1} + \left(\frac{\dot{\h}}{\h} +
\frac{2}{a}\right)\dot{\delta}_{\rm 1} +
\frac{k^{2}}{a^{2}\h^{2}}c_{\rm s1}^{2}\delta_{\rm 1} =
\frac{1}{\h^{2}}\left(\rho_{\rm 1}\delta_{\rm 1} +
\rho_{\rm 2}\delta_{\rm 2}\right)\\ \\
\ddot{\delta}_{\rm 2} + \left(\frac{\dot{\h}}{\h} +
\frac{2}{a}\right)\dot{\delta}_{\rm 2} +
\frac{k^{2}}{a^{2}\h^{2}}c_{\rm s2}^{2}\delta_{\rm 2} =
\frac{1}{\h^{2}}\left(\rho_{\rm 1}\delta_{\rm 1} + \rho_{\rm
2}\delta_{\rm 2}\right).
\end{array}
\right.
\end{equation}
This approximation is sufficient to study the formation of all
gravitationally bound objects (galaxies, in particular) which become
nonlinear for $z > 1$. As pointed out in \cite{SolovStar}, it is
possible to recover the system (\ref{gensyssimplified}) by
considering the limit $k \gg \h$, too, but it is not a necessary
constraint in the matter--dominated regime when $c_{\rm s1}$ and
$c_{\rm s2}$ are small. It is interesting that the system
(\ref{gensyssimplified}) can be exactly solved for the two--fluid
model considered in Sec. \ref{gcgandbar}. The solution is exhibited
in the Appendix.

\section{Perturbations in a Chaplygin gas dominated
universe}\label{gcgruled}

In the gCg dominated universe the Hubble parameter scales as
follows:
\begin{equation}\label{gcgH}
\frac{\mathcal{H}^{2}}{\h_{\rm 0}^{2}} = \left(\bar{A} + \frac{1 -
\bar{A}}{a^{3(1 + \alpha)}}\right)^{\frac{1}{1 + \alpha}}a^{2},
\end{equation}
where $\bar{A} = A/(A+B)$ and $\h_{\rm 0} = H_{\rm 0}$ is the Hubble
constant.

In a typical analysis of the gCg model it is customary to take
$\alpha < 1$ since the square sound velocity tends to $\alpha$ for
$a \to \infty$, as can be seen from (\ref{gcgcs}). For $\alpha = 1$
the sound velocity will tend to unity in the far future, at $t\to
\infty$. The new regime that we investigate in the present paper is
$\alpha > 1$, thereby making gCg superluminal after a sufficiently
long but still finite time.

 From (\ref{gcgcs}) it is straightforward to calculate at which
redshift the transition to the superluminal gCg would occur. For
generic values of $\bar{A}$ and $\alpha$:
\begin{equation}\label{transz}
z_{\rm s} = \left[\frac{\bar{A}\left(\alpha - 1\right)}{1 -
\bar{A}}\right]^{\frac{1}{3\left(\alpha + 1\right)}} - 1.
\end{equation}
Given $\alpha >1$ and $\bar{A}$, it is plausible to expect that
some, hopefully observable, cosmological effect would take place at
that redshift.

It is crucial to point out that $\alpha$ and $\bar{A}$ are not
independent, but linked by (\ref{gcgH}), once we have some
constraints about the background expansion. In the present paper we
consider $z_{\rm tr}$, i.e. the redshift at which the transition to
the accelerated phase of the expansion takes place. We derive its
expression from (\ref{gcgH}):
\begin{equation}\label{acctransz}
z_{\rm tr} = \left[\frac{2\bar{A}}{1 -
\bar{A}}\right]^{\frac{1}{3\left(\alpha + 1\right)}} - 1.
\end{equation}

It follows from the most recent SN Ia observations combined with CMB
and baryon acoustic oscillations (BAO) data that $z_{\rm tr} \approx
0.7$ with a rather large uncertainty (at least $\pm 0.1$ or more at
the $2\sigma$ confidence level), see e.g. Fig. 7 in \cite{ASS07} and
also \cite{Davis,MPP07,Daly07}. For comparison, if only SN Ia are
used, then the Gold+HST dataset \cite{Riess07} leads to $z_{\rm tr}
\sim 0.4$, see Fig. 2 in \cite{ASS07} and also \cite{DD07}) while
other datasets (SNLS and ESSENCE) alone produce values of $z_{\rm
tr}\sim 0.7$, too (see \cite{Davis, ASS07}). However, these numbers
strongly depend on the present value of the non-relativistic matter
density $\Omega_{\rm m0}$ which, in turn, has to be expressed
through $\bar{A}$ in the gCg case (the values given above correspond
to $\Omega_{\rm m0}\approx 0.28$). Direct calculation of $z_{\rm
tr}$ for UDM models \cite{WY07,MPP07} shifts this quantity to even
larger redshifts: $z_{\tr}=0.8-0.9$. Actually, in the latter case,
the authors consider only silence quartessence (namely the effective
sound velocity is assumed to be zero) and the parameter $\alpha$ is
constrained to be $\alpha = -0.06 \pm 0.1$.

Then from (\ref{acctransz}) we can write down an explicit relation
between $\bar{A}$ and $\alpha$:
\begin{equation}\label{acctranszA}
\bar{A} = \frac{\left(1 + z_{\rm tr}\right)^{3\left(1 +
\alpha\right)}}{2 + \left(1 + z_{\rm tr}\right)^{3\left(1 +
\alpha\right)}}.
\end{equation}
Note that the uncertainty $\sigma_{z_{tr}}$ of the transition
redshift propagates on $\bar{A}$ through the following formula:
\begin{equation}\label{acctranszAerr}
\sigma_{\bar{A}} = \frac{6(\alpha+1)}{[2(1 + z_{\rm
tr})^{-3(\alpha+1)} + 1]^{2}(1 + z_{\rm
tr})^{3\alpha+4}}\sigma_{z_{tr}},
\end{equation}
therefore the larger is $\alpha$, the smaller is the bias on
$\bar{A}$.

A widely used cosmological parameter stemming from CMB observation
is $R$, the comoving distance to the last scattering surface scaled
to $\Omega_{\rm m0}$, namely:
\begin{equation}\label{Rparameter}
R = H_{\rm 0}\sqrt{\Omega_{\rm m0}}\int_{0}^{z_{\rm
ls}}\frac{dz}{H(z)},
\end{equation}
where $z_{\rm ls} \approx 1089$ is the last scattering surface
redshift. In UDM models, the quantity $\Omega_{\rm m0}$ should be
taken from the asymptotic behaviour of ${\cal H}^2$ for $z\gg 1$ at
the matter--dominated stage, so $\Omega_{\rm m0} = \left(1 -
\bar{A}\right)^{1/(\alpha + 1)}$ (or $\Omega_{\rm m0} =
\left(1-\bar{A}\right)^{1/(\alpha+1)}(1-\Omega_b)+\Omega_b$ if
baryons are taken into account, too). At the $1\sigma$ confidence
level, $R = 1.71 \pm 0.03$ (see for example \cite{Wang07}).

An interesting limiting case is the 'super--duperluminal' one:
$\alpha\to\infty$. Then $H\equiv {\cal H}/a={\rm const}=H_0$ for
$z\le z_{\rm tr}$ and $H^2(z)=H_0^2 \left((1+z)/(1+z_{\rm
tr})\right)^3$ for larger $z$. In this case, one can obtain an
analytic expression for $R$:
\begin{equation} \label{Rinfty}
R_{\infty}=\frac {z_{\rm tr}} {(1+z_{\rm tr})^{3/2}} + \frac
{2}{\sqrt{1+z_{\rm tr}}} - \frac {2}{\sqrt{1+z_{\rm ls}}}~.
\end{equation}
Then the above mentioned observational window for $R$ is reached for
$z_{\rm tr}=1.0\pm 0.1$.

Combining (\ref{acctransz}) with (\ref{transz}) we find that
\begin{equation}\label{acctransz2}
z_{\rm s} = \left(\frac{\alpha -
1}{2}\right)^{\frac{1}{3\left(\alpha + 1\right)}}\left(1 + z_{\rm
tr}\right) - 1.
\end{equation}
If $\alpha = 3$, then $z_{\rm s} = z_{\rm tr}$. For $\alpha > 3$,
$z_{\rm s} > z_{\rm tr}$ but $z_{\rm s}$ approaches $z_{\rm tr}$
once more for $\alpha\to\infty$. For a fixed $z_{\rm tr}$, the
maximal value of $z_{\rm s}$ is reached for $\alpha\approx 8.182$
when $(1+z_{\rm s})/(1+z_{\rm tr})\approx 1.048$. Thus, for $\alpha
\gtrsim 3$, the transition from subluminal to superluminal gCg
occurs approximately at the same time as the transition to the
accelerated phase of expansion of the Universe.

We now compare the evolution of perturbations in the gCg--dominated
universe for different values of $\alpha$. We choose an integration
range which starts at the decoupling era, namely $z \sim 1100$ ($a
\approx 10^{-3}$), and ends up when the first structures, namely
protogalaxies, were formed ($a \approx 0.1$, or $z \approx 10$), so
that the linear approximation holds true.

We numerically solve (\ref{eqPhi}) choosing $\left[-1, 0\right]$ as
normalized initial conditions for $\left[\Phi, \dot{\Phi}\right]$.
In our calculations we use (\ref{acctranszA}) to properly choose
$\bar{A}$ as a function of $\alpha$. We show the results using
$z_{tr} = 0.8 \pm 0.16$ and $z_{tr} = 1.0 \pm 0.1$.
\begin{figure}[htbp]
\includegraphics[width=13cm]{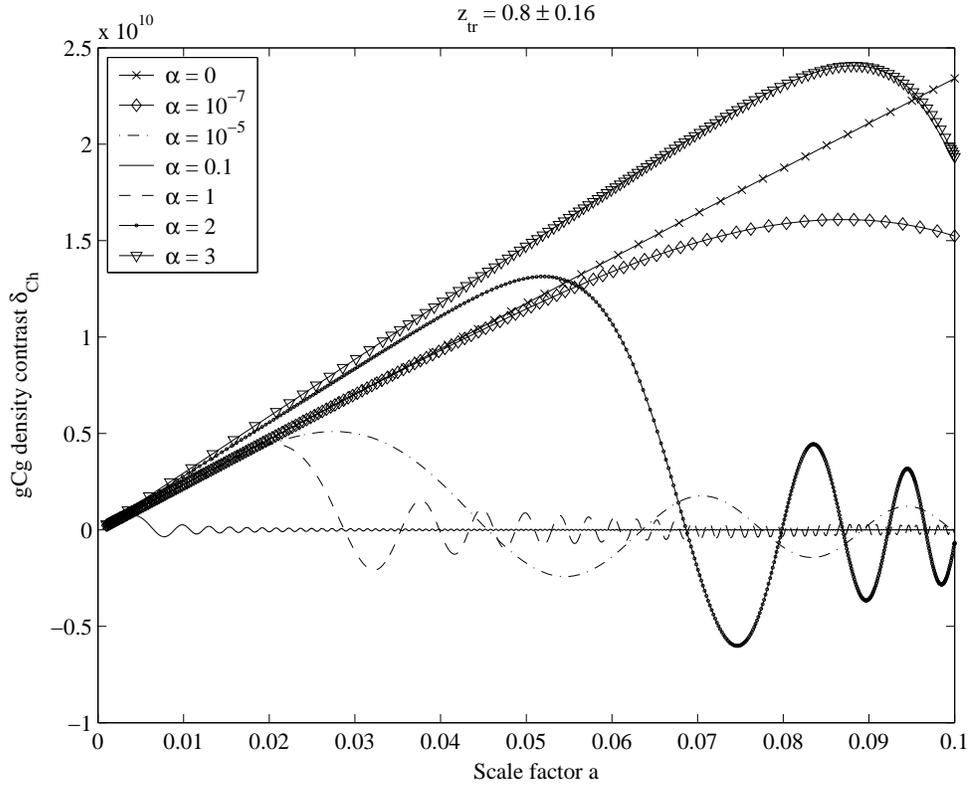}\\
  \caption{Evolution profiles of $\delta_{\rm Ch}$ for different
values of $\alpha$
  and for $k = 100$ h Mpc$^{-1}$. The transition redshift is $z_{tr}
= 0.8 \pm 0.16$}\label{Fig1}
\end{figure}
\newpage
\begin{figure}[htbp]
\includegraphics[width=13cm]{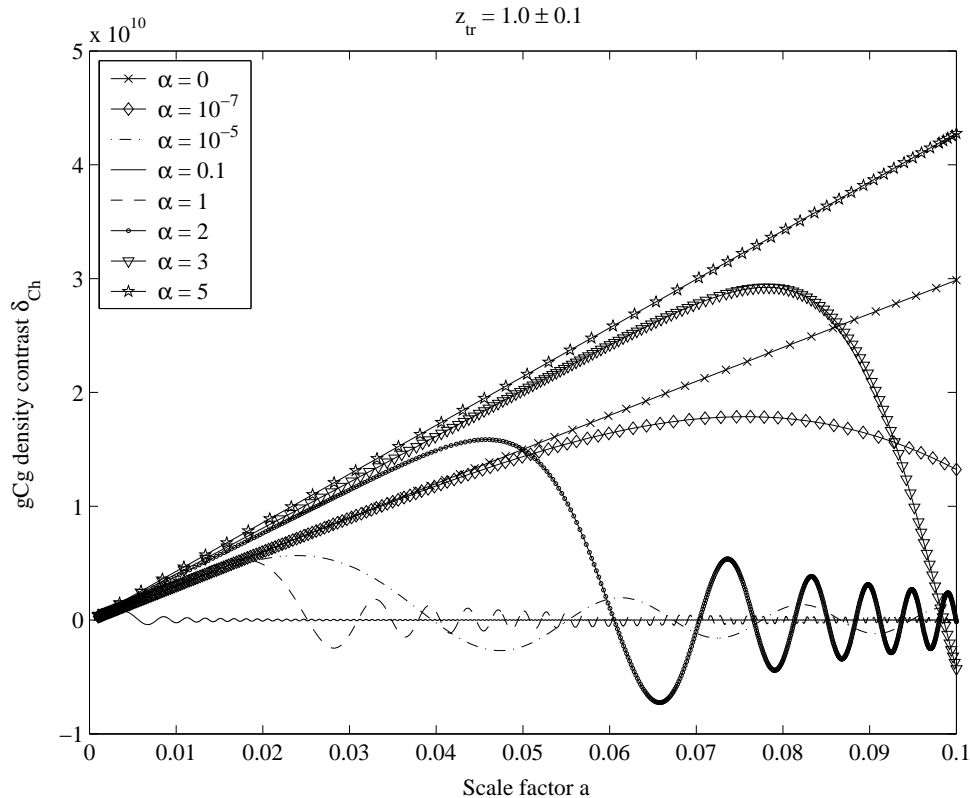}\\
  \caption{Same as figure \ref{Fig1}, with $z_{tr} = 1.0
\pm 0.1$}\label{Fig2}
\end{figure}
The plots in figures \ref{Fig1} and \ref{Fig2} display the evolution
of the density contrast in the gCg, which we call $\delta_{\rm Ch}$,
for different choices of the parameter $\alpha$ and for $k = 100$ h
Mpc$^{-1}$, which corresponds to a scale of order $50$ $h^{-1}$
$kpc$, which is typical of a protogalaxy. The case $\alpha = 0$
corresponds to the $\Lambda$CDM model.

Indeed, here we are first investigating if formation of
gravitationally bound objects including galaxies takes place at all.
That is why we have chosen so small a scale and we have stopped at
$z = 10$. The gCg UDM model, like all CDM--like ones, belongs to the
class of bottom--up, or hierarchical clustering, scenarios of
structure formation where small scale structures are formed first.
The question of large--scale structure in this model at scales which
remain linear for $z < 1$, of course, requires integration up to $z
= 0$. It is considered in the next section.

Part of the results found in \cite{Sandvik} are already confirmed by
this analysis, without a deeper study involving the power spectrum.
The development of oscillations takes place too early for $\alpha
\gtrsim 10^{-5}$, thus preventing structure formation.

However, two further interesting features can be extracted from
figures \ref{Fig1} and \ref{Fig2}. First, it seems that there exists
a critical value, namely $\alpha \approx 0.1$, for which the
deviation of the gCg from the $\alpha = 0$ linear growth is maximal.
Second, the larger the value of $\alpha$ (in the range $\alpha >
0.1$) the smaller the deviation from the $\alpha = 0$ behaviour.
Indeed, when $\alpha > 1$, i.e. in the superluminal regime, the
oscillations are absent and the growth seems even to be enhanced.

\begin{figure}[htbp]
  \includegraphics[width=13cm]{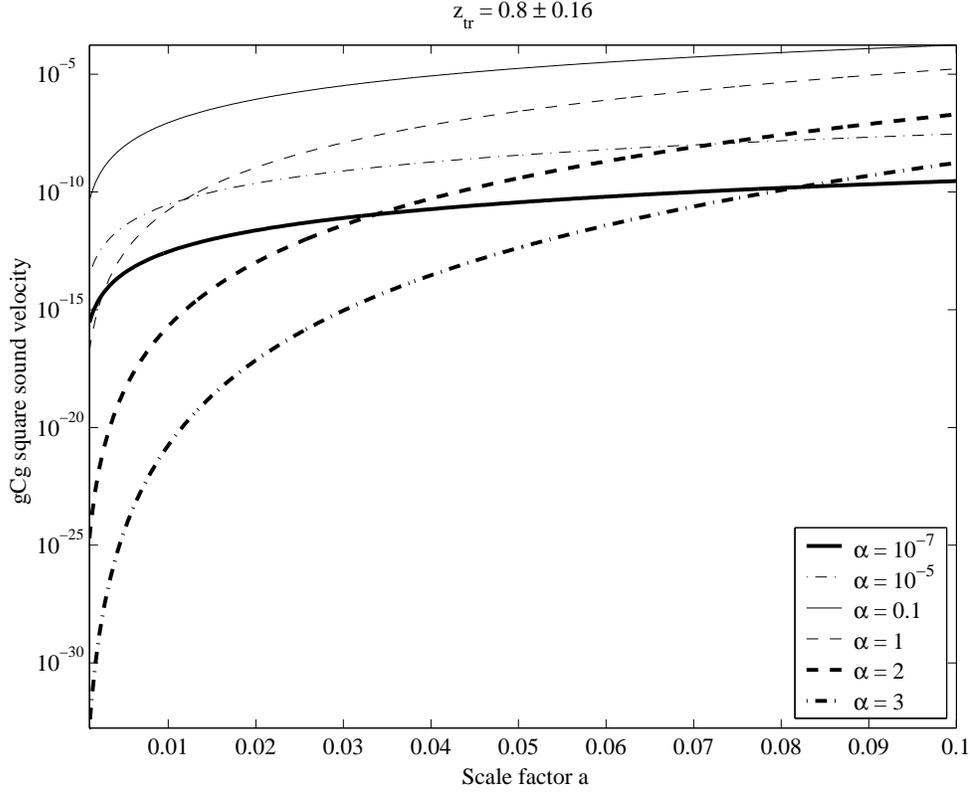}\\
  \caption{The gCg square sound velocity given in  (\ref{gcgcs}) plotted
as a function of $a$ for different
  values of $\alpha$. $\bar{A}$ is given by  (\ref{acctranszA}) with
$z_{tr} = 0.8 \pm 0.16$.}\label{Fig3}
\end{figure}

Both the above mentioned features can be explained by the behaviour
of the gCg sound velocity as a function of $\alpha$, which is
displayed in figure \ref{Fig3} for $z_{tr} = 0.8 \pm 0.16$ (the
plots for $z_{tr} = 1.0 \pm 0.1$ are very similar).
\newline
For $\alpha \sim 0.1$ the sound velocity becomes non negligible much
earlier than at other values of $\alpha$. The range $10^{-7}
\lesssim \alpha \lesssim 3$ appears  thus to be ruled out since
structure formation is prevented. However, to understand what
happens out of this range it is necessary to study the clustering
properties of matter at the present time, namely the power spectrum.
This will be done in the following section.

\newpage

\section{Large Scale Power spectrum of Chaplygin gas
perturbations}\label{gcgpowers}

In the previous section we have shown that in the gCg model galaxy
formation is possible only when $\alpha$ is very small or $\alpha
\gtrsim 3$. In this section we try and constrain these bounds
further by studying the clustering properties of gCg perturbations,
namely the power spectrum $P(k)$, and comparing it with observation.

We exploit the SDSS (Sloan Digital Sky Survey) data as analysed in
\cite{Teg3}. The sample considered by the authors consists of
205,443 galaxies observed before July 2002. The data consist of RA
(Right Ascension), Dec (Declination) and redshift $z$ for each
galaxy. The redshift is converted to comoving distance by using a
flat cosmological model with a cosmological constant $\Omega_{\rm
\Lambda} = 0.7$ (it can be shown that our results are robust to this
assumption).

The power spectrum is computed at present time, namely at $z = 0$.
The data processed in \cite{Teg3} are displayed in \tref{tabPk}.
\begin{table}[htbp]
\caption{\label{tabPk} Effective $k$, window functions and measured
values of $P(k)$ with relative bias factors and standard deviations.
$[k] = h Mpc^{-1}$ and $\left[P(k)\right] = (h^{-1} Mpc)^{3}$. From
the SDSS data reported in \cite{Teg3}. We have neglected two rows of
data in which $\sigma$ is greater than the respective $P(k)$.}
\lineup
\begin{indented}
\item[]\begin{tabular}{@{}lllllll} \br
  \textbf{k} & \textbf{Low k} & \textbf{High k} &
\textbf{$P(k)$} & \textbf{Bias}
  & \textbf{$\sigma$}\\
  \mr
     0.01819  &    0.01502   &  0.02440  &  33254.63896 & 1.16745 &
     24572.79357\\
     0.02405    &  0.01984   &   0.03136 & 38360.58264 &  1.16674 &
     13320.48227\\
     0.02776    &  0.02300   &   0.03571 & 24143.07699 & 1.16613 &
     10047.21709\\
     0.03201    &  0.02663   &   0.04048 & 19709.29306 & 1.16532 &
     \07413.82995\\
     0.03691    &  0.03094   &   0.04601 & 12595.77528 & 1.16428 &
     \05485.84481\\
     0.04257    &  0.03591   &   0.05215 & 13558.60084 & 1.16294 &
     \04077.68223\\
     0.04912    &  0.04162   &   0.05983 & 18311.08054 & 1.16124 &
     \02974.15048\\
     0.05670    &  0.04824   &   0.06873 & 12080.73574 & 1.15910 &
     \02140.39514\\
     0.06527    &  0.05561   &   0.07867 &  \09217.46084 & 1.15647 &
     \01580.01728\\
     0.07529    &  0.06455   &   0.08999 &  \09750.49986 & 1.15317 &
     \01127.95803\\
     0.08698    & 0.07507    &  0.10351  &  \09529.63935  & 1.14906 &
     \0\0818.49019\\
     0.10037    &  0.08670   &   0.11898 &  \06384.82545 & 1.14409 &
     \0\0601.70752\\
     0.11581     & 0.09985   &   0.13679 &  \05294.92081 & 1.13813 &
     \0\0446.65015\\
     0.13360     & 0.11478   &   0.15748 &  \04629.83669 & 1.13109 &
     \0\0335.07950\\
     0.15412    & 0.13181    &  0.18139  &  \03573.92724  & 1.12290 &
     \0\0254.13547\\
     0.17774    &  0.15114   &   0.20891 &  \03393.82814 & 1.11358 &
     \0\0195.23431\\
     0.20489    &  0.17288   &   0.24062 &  \02298.10000 & 1.10320 &
     \0\0152.73819\\
     0.23592    &  0.19690   &   0.27653 &  \01597.14976 & 1.09193 &
     \0\0123.50981\\
     0.27062    &  0.22219   &   0.31395 &  \01105.42903 & 1.08019 &
     \0\0107.20069\\
     0.30618    &  0.23123   &   0.34782 &  \01012.72489 & 1.06912 &
     \0\0110.33632\\
     \br
\end{tabular}
\end{indented}
\end{table}
\newline
We compute the gCg power spectrum for different values of $\alpha$,
by first solving (\ref{eqPhi}), thus finding the transfer function
and then applying it to the following prior:
\begin{equation}\label{BBKS}
\fl \left|\delta_{k}\right|^{2} = Nk\left[\frac{\ln\left(1 +
2.34q\right)}{2.34q}\right]^{2}\left[1 + 3.89q +
\left(16.1q\right)^{2} + \left(5.46q\right)^{3} +
\left(6.71q\right)^{4}\right]^{-1/2},
\end{equation}
where $N$ is a normalization constant and
\begin{equation}\label{q}
q \equiv \frac{k}{\Omega_{X0}h^{2}},
\end{equation}
where $\Omega_{X0}$ is the cold dark matter (CDM) energy density
fraction evaluated today, and $h$ is the Hubble constant in units of
$100$ $km$ $s^{-1}$ $Mpc^{-1}$. In our calculations we shall set $h
= 0.7$. Since the gCg has the same asymptotic behaviour as the
$\Lambda$CDM model in the matter--dominated stage, then the cold
dark matter fraction is given by
\begin{equation}\label{shapeparameter}
\Omega_{X0} = \left(1 - \bar{A}\right)^{\frac{1}{1 + \alpha}}.
\end{equation}

The prior (\ref{BBKS}) is obtained by applying the BBKS transfer
function \cite{Bardeen2} to the scale invariant Harrison--Zel'dovich
spectrum. The normalization constant $N$ is computed through a best
fit of the data for each $\alpha$. In our calculation, $\bar{A}$ is
constrained by (\ref{acctranszA}) using $z_{tr} = 0.8 \pm 0.16$ and
$z_{tr} = 1.0 \pm 0.1$.
\begin{figure}[htbp]
  \includegraphics[width=13cm]{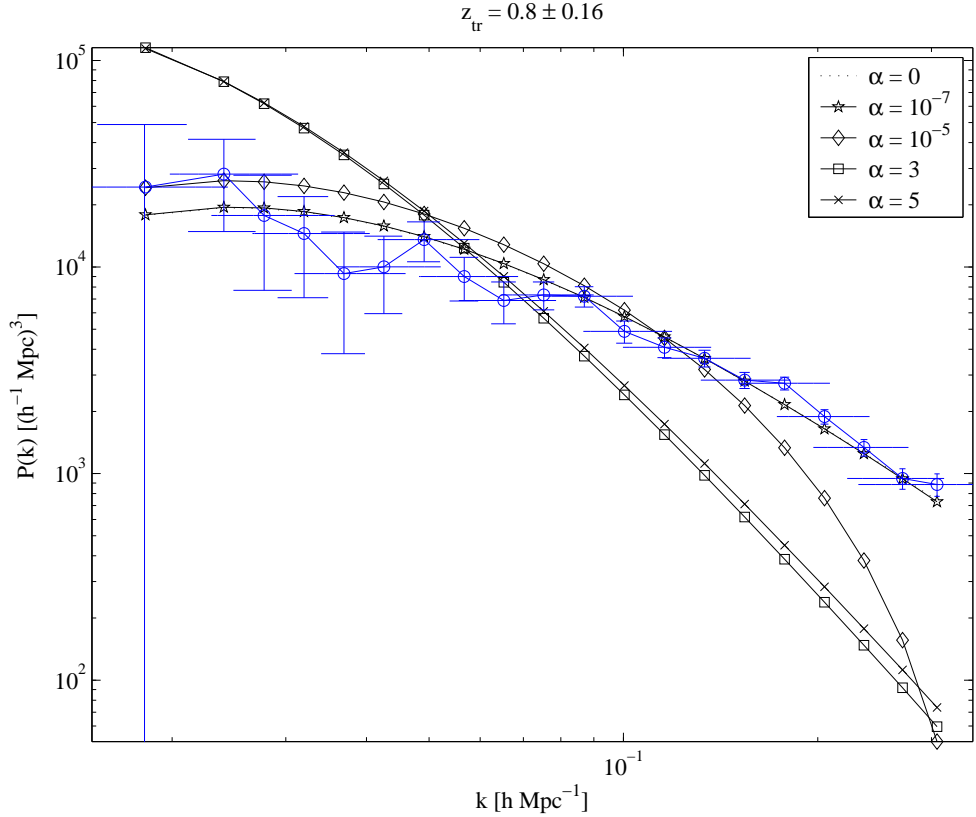}\\
  \caption{Theoretical power spectra for different values of $\alpha$ compared
  with the measured one. $\bar{A}$ is given by (\ref{acctranszA}) with $z_{tr} = 0.8 \pm 0.16$. Note that the plots
  for $\alpha = 0$ and $\alpha = 10^{-7}$ are superposed. We have
  extracted the envelopes of the oscillations of the density contrasts for $\alpha = 3$ and $\alpha = 5$
  and have used them for the calculations. For the other cases this procedure was unnecessary since no
  oscillations were present.}\label{Fig4}
\end{figure}
\newpage
\begin{figure}[htbp]
  \includegraphics[width=13cm]{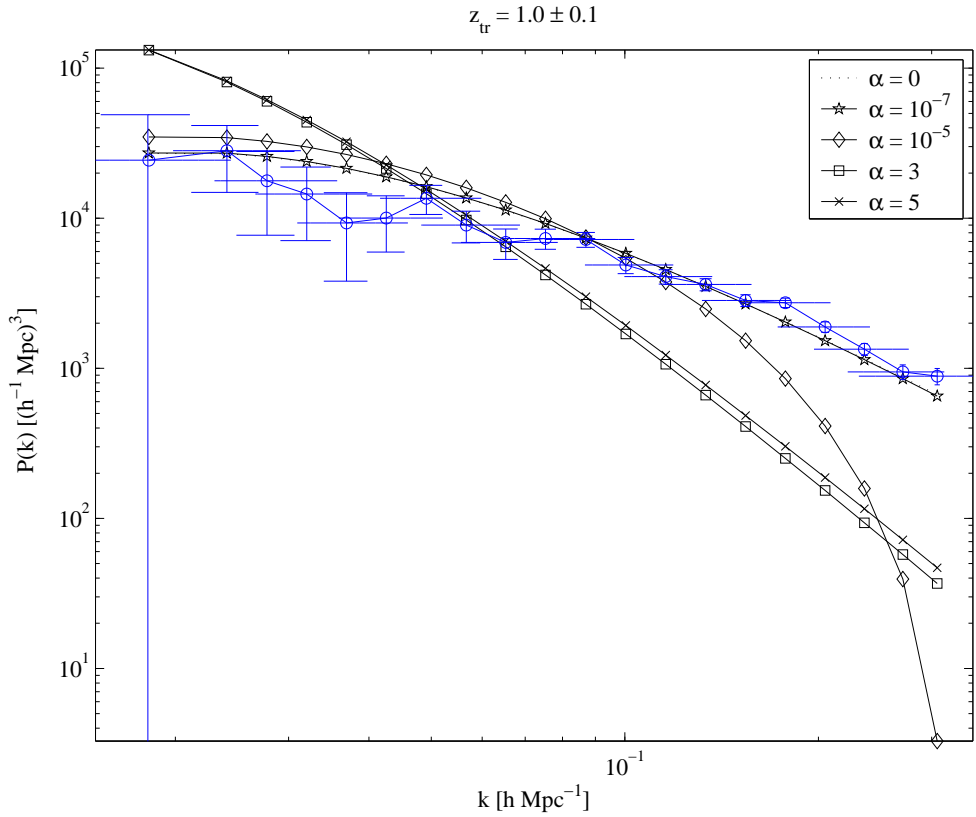}\\
  \caption{Same as figure \ref{Fig4}, with $z_{tr} = 1.0 \pm 0.1$.}\label{Fig5}
\end{figure}
In figures \ref{Fig4} and \ref{Fig5} the power spectra have been
computed by considering the envelope of the oscillations of the
density contrast $\delta_{Ch}$ and by using it in the subsequent
calculations. This procedure serves to clearly compare the different
slopes of the calculated power spectra with respect to the observed
one. Moreover, it has been carried out only for $\alpha = 3$ and
$\alpha = 5$. For the other ones it was unnecessary as no
oscillations were present.

An interesting feature that can be drawn from figures \ref{Fig4} and
\ref{Fig5} is that the larger is $\alpha$ the more the power
spectrum tends to a limiting behaviour which is systematically below
that of the $\Lambda$CDM one. This happens because the suppression
of the gCg transfer function relative to that of the $\Lambda$CDM
model behaves approximately as $k^{-1}$ for large values of $\alpha$
after averaging over oscillations (actually, the exponent is
slightly larger than $-1$ because of the accelerated expansion at
recent times).
\begin{figure}[htbp]
  \includegraphics[width=13cm]{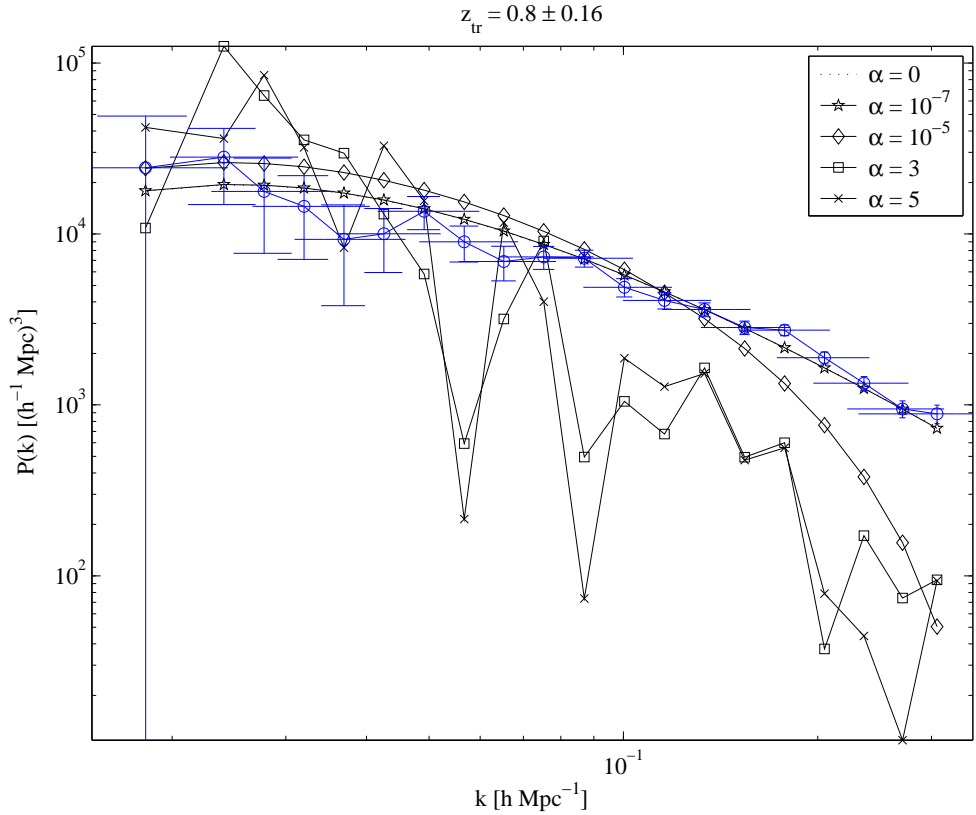}\\
  \caption{Theoretical power spectra for different values of $\alpha$ compared
  with the measured one. $\bar{A}$ is given by (\ref{acctranszA}) with $z_{tr} = 0.8 \pm 0.16$. Note that the plots
  for $\alpha = 0$ and $\alpha = 10^{-7}$ are superposed.}\label{Fig4bis}
\end{figure}
\begin{figure}[htbp]
  \includegraphics[width=13cm]{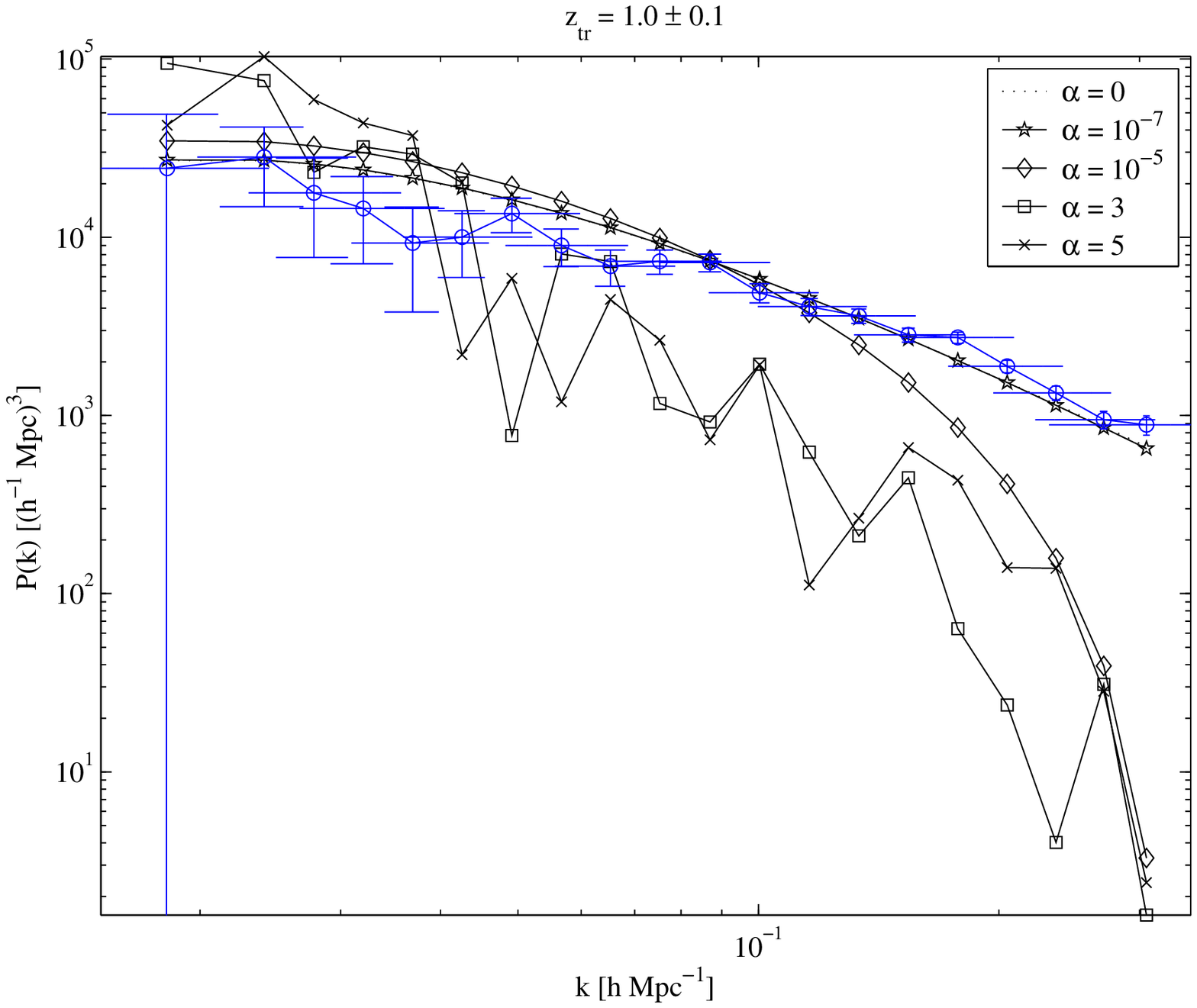}\\
  \caption{Same as figure \ref{Fig4bis}, with $z_{tr} = 1.0 \pm 0.1$.}\label{Fig5bis}
\end{figure}

Figures \ref{Fig4bis} and \ref{Fig5bis} show what happens if the
oscillations in the density contrasts are taken into account. The
power spectra for $\alpha = 3$ and $\alpha = 5$ are highly irregular
and their slopes are hardly recognizable.

\newpage

In conclusion, the pure gCg does not seem to work from a
perturbative viewpoint, even when it is superluminal. However, a
reasonable possibility is to allow the gCg to have non--adiabatic
perturbations, which is a natural assumption since it is not a
pressureless fluid. An attempt in this direction has already been
performed in \cite{Reis} and \cite{Amendola} (here also a baryon
component is considered). However, the results are based on the {\it
ad hoc} assumption of silent perturbations (namely $\delta p_{\rm
Ch} = 0$), which seems questionable.

\newpage

\section{Two-fluid model: generalized Chaplygin gas plus
baryons}\label{gcgandbar} In this section we consider a two fluid
model including baryons together with gCg. The Friedmann equation
governing this model has the following form:
\begin{equation}\label{chbHOmega}
\frac{\mathcal{H}^{2}}{\h_{\rm 0}^{2}} = \left\{\frac{\Omega_{\rm
b0}}{a^{3}} + \left(1 - \Omega_{\rm b0}\right)\left[\bar{A} +
\frac{1 - \bar{A}}{a^{3\left(\alpha + 1\right)}}\right]^{1/1 +
\alpha}\right\}a^{2},
\end{equation}
where subscript {\rm b} refers to the baryonic component and
subscript $0$ indicates that the corresponding quantity is evaluated
at the present epoch ($a = 1$).

System (\ref{gensys}) is solved numerically in the same integration
range of the previous sections and setting $\Omega_{\rm b0} = 0.04$.
Before proceeding it is necessary  to find a relation between
$\bar{A}$ and $\alpha$ which ensures the preservation of the correct
background expansion, e.g. the acceleration redshift $z_{\rm tr}
\approx 0.6$.

From (\ref{chbHOmega}) it is possible to work out the following
expression:
\begin{equation}\label{acctranszAChb}
\frac{\bar{A}\left[(1+z_{\rm tr})^{-3(\alpha+1)} +
1/2\right]-1/2}{\left\{\bar{A}\left[(1+z_{\rm
tr})^{-3(\alpha+1)}-1\right] + 1\right\}^{\alpha/\alpha + 1}} =
\frac{\Omega_{\rm b0}}{2\left(1-\Omega_{\rm b0}\right)},
\end{equation}
which cannot be inverted to give $\bar{A}$ as a function of
$\alpha$. However, since $\Omega_{\rm b0}$, is small we can use
again (\ref{acctranszA}) to that purpose.

As in the previous section, we choose (\ref{BBKS}) as a prior for
the power spectrum, but since this time we have the presence of
baryons, $\Omega_{X0}$ has to be chosen differently. We exploit then
Sugiyama's shape correction \cite{Sugiyama}:
\begin{eqnarray}\label{Sugi}
\fl \Omega_{X0} = \left[\Omega_{b0} + \left(1 -
\Omega_{b0}\right)\left(1 - \bar{A}\right)^{\frac{1}{1 +
\alpha}}\right]\times \nonumber\\
\exp\left(-\Omega_{b0} - \frac{\sqrt{2h}\Omega_{b0}}{\Omega_{b0} +
\left(1 - \Omega_{b0}\right)\left(1 - \bar{A}\right)^{\frac{1}{1 +
\alpha}}}\right),
\end{eqnarray}
which properly takes into account the presence of baryons.

In the appendix we show an exact solution, in the matter dominated
regime, for the evolution of perturbations in the model of the
present section.
\begin{figure}[htbp]
  \includegraphics[width=13cm]{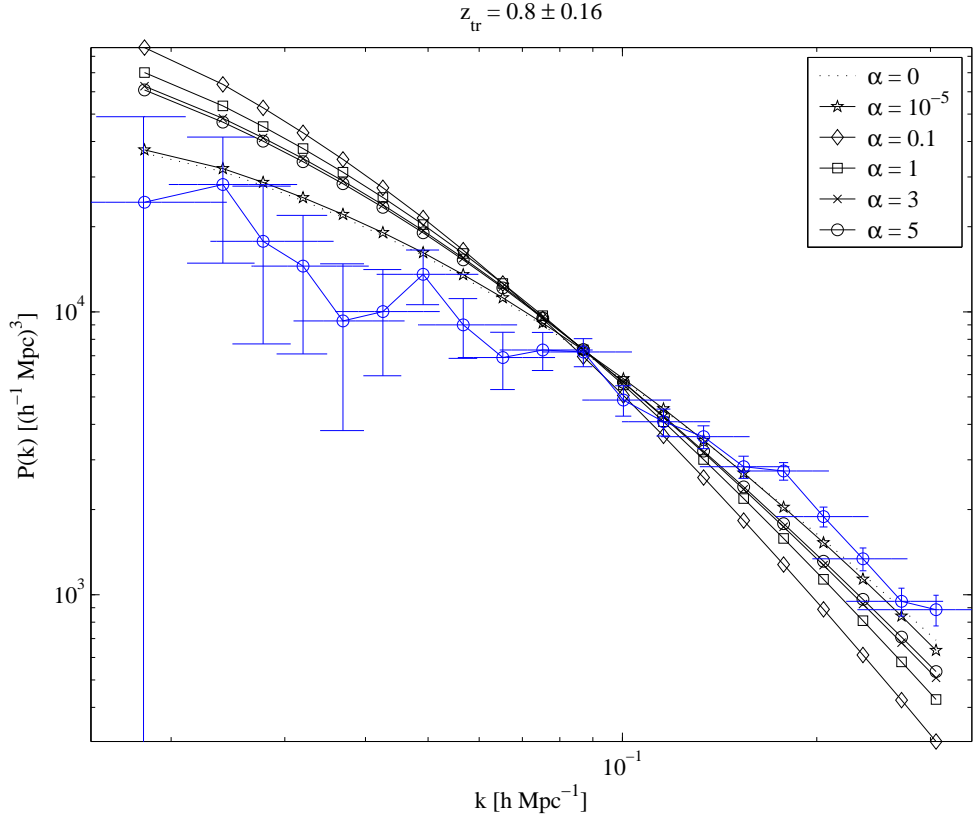}\\
  \caption{Power spectra of the baryonic component computed for different values of $\alpha$ with $z_{tr}
= 0.8 \pm 0.16$.}\label{Fig6}
\end{figure}
\newpage
\begin{figure}[htbp]
  \includegraphics[width=13cm]{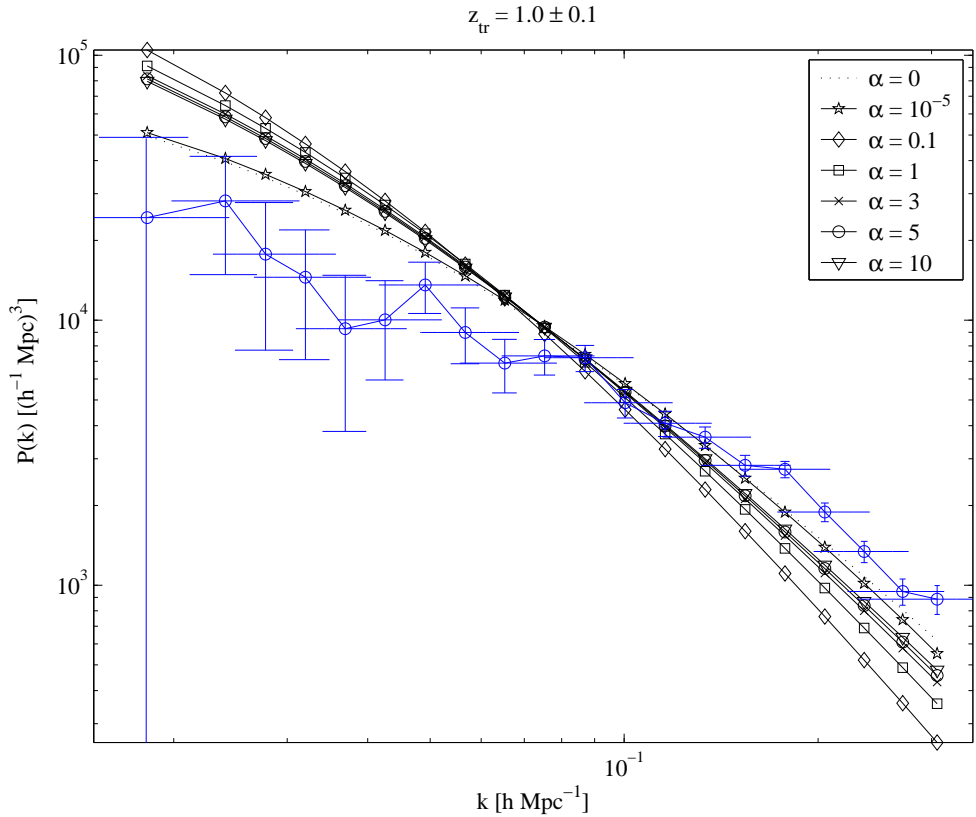}\\
  \caption{Same as figure \ref{Fig6} with $z_{tr}
= 1.0 \pm 0.1$.}\label{Fig7}
\end{figure}
In figures \ref{Fig6} and \ref{Fig7} we show the baryonic power
spectra computed for different values of $\alpha$ compared to the
observed one. We have chosen as normalized initial conditions the
following values: $\left[\delta_{\rm b}, \dot{\delta}_{\rm b},
\delta_{\rm Ch}, \dot{\delta}_{\rm Ch}\right] = \left[1, 1, 1,
1\right]$. We have chosen the same initial conditions for all the
quantities essentially because after the equivalence era the gCg
behaved as dust and therefore its density contrast grew linearly (as
a function of the scale factor).

Unlike what happened in the previous section, this time there were
no oscillations in $\delta_{b}$, since the baryonic density contrast
is unaffected by the instabilities of the gCg one. In fact, the
results are different from those inferred from figures \ref{Fig4}
and \ref{Fig5}. The agreement with the observed power spectrum is
better. It is particularly good again for very small $\alpha$ but
also in the range $\alpha > 1$. Therefore the gCg seems to work
better either when it is practically indistinguishable from
$\Lambda$CDM or when its sound velocity is superluminal.

\begin{figure}[htbp]
  \includegraphics[width=13cm]{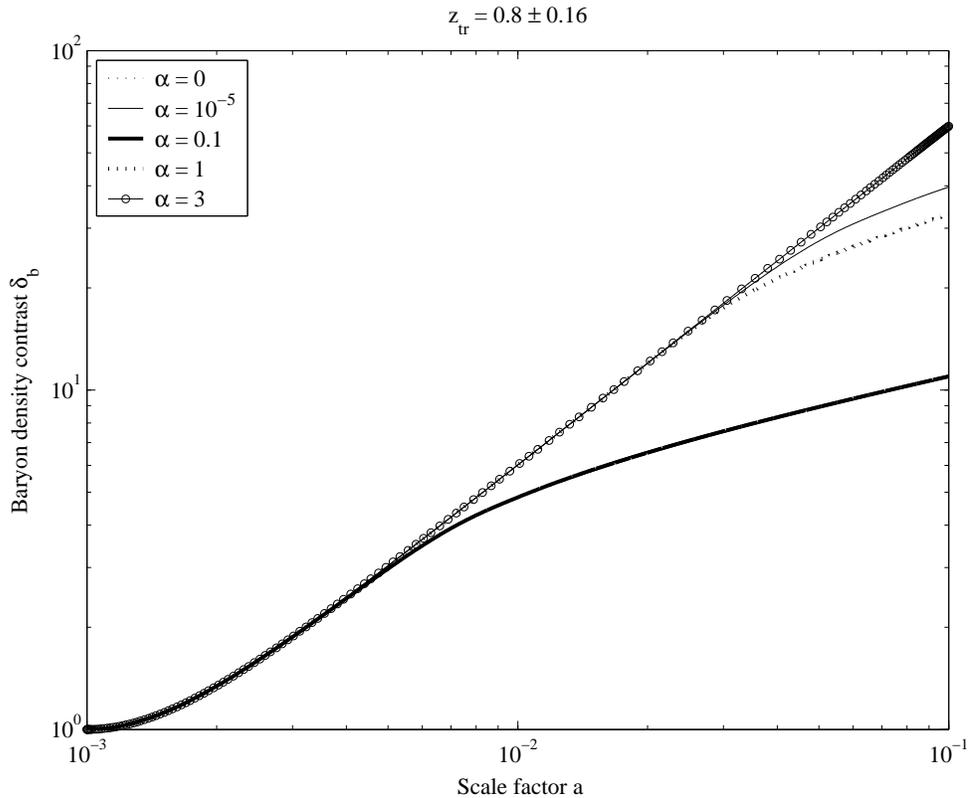}\\
  \caption{Evolution profiles of $\delta_{\rm b}$ for different values of $\alpha$
  and for $k = 100$ $h$ $Mpc^{-1}$. Note that the plot for $\alpha = 0$ and $\alpha = 3$
  are superposed. $z_{tr}
= 0.8 \pm 0.16$.}\label{Fig8}
\end{figure}

\begin{figure}[htbp]
  \includegraphics[width=13cm]{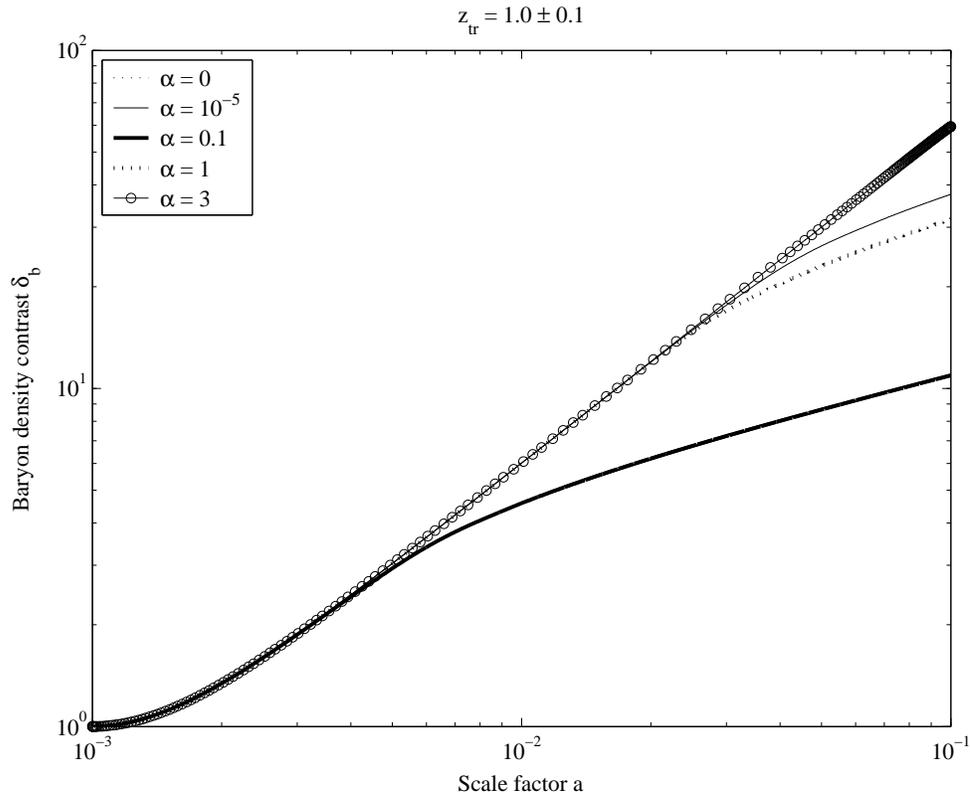}\\
  \caption{Same as figure \ref{Fig8}, with $z_{tr}
= 1.0 \pm 0.1$.}\label{Fig9}
\end{figure}

The good agreement of the theoretical power spectrum with the
observed one can be explained realizing that gCg perturbations are
indistinguishable from those of baryons at the beginning of the
matter--dominated stage. Therefore they really act as dark matter
and compel baryons perturbations to grow and form structures. On the
other hand, when the gCg sound velocity increases, the growth of the
baryons inhomogeneities is damped, but does not oscillate. This fact
can be observed in figures \ref{Fig8} and \ref{Fig9}. Moreover,
structures which have already formed are not influenced by the gCg
sound velocity, so neither is the large scale power spectrum. This
was already pointed out in \cite{Beca}.
\newline
In figures \ref{Fig8} and \ref{Fig9} we display the evolution of
$\delta_{\rm b}$ for different values of $\alpha$ choosing again
$\left[\delta_{\rm b}, \dot{\delta}_{\rm b}, \delta_{\rm Ch},
\dot{\delta}_{\rm Ch}\right] = \left[1, 1, 1, 1\right]$ as
normalized initial conditions. As in section \ref{gcgruled} we have
chosen $k = 100$ $h$ $Mpc^{-1}$. Note that the plots corresponding
to $\alpha = 0$ and $\alpha = 3$ are superposed.

We infer from figures \ref{Fig8} and \ref{Fig9} that the gCg sound
velocity has a damping effect on the growth of baryon
inhomogeneities. These are not compelled to oscillate, but structure
formation is delayed, if not prevented. In order to render structure
formation as similar as possible to that of $\Lambda$CDM, we need
either very small values of $\alpha$ or $\alpha > 3$. The latter
range for $\alpha$ just leads to $z_{\rm s} > z_{\rm tr}$ as was
shown in section \ref{gcgruled}.

\newpage

\section{Conclusions}\label{concl}

From our analysis it emerges that:
\begin{itemize}
  \item The generalized Chaplygin gas cosmological model, with no
  additional fluid components, is compatible with
  structure formation and large scale structure clustering
  properties (encoded in the power spectrum) only for $\alpha$
  sufficiently small ($\alpha < 10^{-5}$), in which case it is
  practically indistinguishable from the $\Lambda$CDM model.
  \item Adding to the generalized Chaplygin gas model a baryon
  component we find that the growth of the density
  contrast and the large scale structure clustering properties are
  compatible with observations for all values of $\alpha$. However
  very small values of $\alpha$ and the case $\alpha \gtrsim 3$ are
  favoured.
\end{itemize}

Note, however, that the integrated Sachs-Wolfe (ISW) effect (not
considered in this paper) leads to rather large enhancement of low
CMB multipoles $C_l$ in UDM models, in contrast to a modest increase
of $C_l$ in the $\Lambda$CDM model \cite{KS85}, that results in the
exclusion of the gCg with $0.01< \alpha \le 1$
\cite{Bean,CF03,AFBC03} (see also the recent paper \cite{BB07} for
an analytical treatment of the ISW effect for UDM models). It may be
interesting to investigate if it remains so large for all $\alpha >
1$.

Now, since most of the new results of this paper refer to the
superluminal case $\alpha > 1$, it is natural to consider the
question if this range is physically admissible in more detail. The
fact that the group velocity $c_{\rm sCh}$ in media may exceed the
light velocity in vacuum (unity in our notations) is not, in itself,
unphysical. Not only it is not prohibited from the theoretical point
of view, but this phenomenon has been actually observed in
laboratories in case of light propagation through dispersive media.
As was proved long ago (see e.g. \cite{Brill}), what is really
necessary in order not to violate causality is that the signal (or,
the wavefront) velocity should not exceed $1$.

Then, what is the signal velocity for the gCg? It does not appear
possible to answer this question at the hydrodynamic level since the
gCg phenomenological equation of state in the normal (non-phantom)
case requires $\rho_{\rm Ch}\ge A^{1/(1+\alpha)}$, so that the limit
$\rho_{\rm Ch}\to 0$ corresponding to a wavefront expanding into
vacuum cannot be studied. Therefore, to determine the value of the
signal velocity for the gCg, one has to use some underlying
microscopic field--theoretical model from which the equation of
state (\ref{gcgeos}) arises in macroscopic hydrodynamics.

We will use the so-called tachyon representation of the gCg for this
purpose where the gCg is described by a scalar field $\phi$
($k$-essence) with the Lagrangian density \cite{bentobertosen}
\begin{eqnarray}\label{tach}
{\cal L}\equiv p_{\rm Ch}=-V_0\left[ 1-\left(\frac {X}{V_0}\right)^
{\frac {1+\alpha}{2\alpha}}\right]^{\frac {\alpha}{1+\alpha}}~, ~~
X= \frac {1}{2} \phi_{,\mu}\phi^{,\mu} < V_0~, \\
\rho_{\rm Ch}=2X\frac {d{\cal L}}{dX} - {\cal L} = V_0 \left[
1-\left(\frac {X}{V_0}\right)^{\frac {1+\alpha}{2\alpha}}
\right]^{-\frac {1}{1+\alpha}}~, ~~~A=V_0^{1+\alpha}~.
\end{eqnarray}
Though other microscopic models are also possible (see e.g.
discussion in \cite{GKMPS05} in the case $\alpha = 1$), the present
one has the following advantages: a) its solutions are in
one--to--one correspondence to those of the hydrodynamic gCg if
$X>0$, and b) it has the same equations for perturbations, too.

Now we are able to consider the limit $X\to 0$ corresponding to a
wavefront expanding into vacuum with no $k$-essence in front of it.
To prove that the signal velocity is equal to unity in this case, it
is sufficient to show that the wave equation corresponding to
(\ref{tach}), namely
\begin{equation} \label{tach-we}
{\cal L}_X\phi_{;\mu}^{;\mu} + {\cal L}_{XX}\phi^{,\mu}\phi^{,\nu}
\phi_{;\mu;\nu} = 0,
\end{equation}
has plane-wave solutions of the type $f(x-t)$ in flat space-time
(since the signal velocity is also the infinite momentum limit of
the phase velocity of a wave). It is straightforward to check that
the sufficient condition for this to hold is that both ${\cal L}_X$
and ${\cal L}_{XX}$ are finite at $X=0$ (a similar remark is made in
the recent paper \cite{Vik}, see also \cite{KVW07}). In addition,
${\cal L}_X(X=0)$ should be non-negative in order for ghosts not to
be present.

In the superluminal case $\alpha > 1$, the Lagrangian density
(\ref{tach}) does not satisfy this condition since ${\cal L}_X$ and
${\cal L}_{XX}$ diverge at $X\to 0$. However, without going into a
more detailed analysis of the singularity structure at a null
hyper--surface in this case, we propose a simple {\em sufficient}
way to cure the causality problem for the gCg. Namely, let us assume
that the expression (\ref{tach}) changes to
\begin{equation} \label{change}
{\cal L} = -V_0 +  X\left(\frac {V_0}{V_1}\right)^{\frac {\alpha
-1}{2\alpha}}
\end{equation}
for $X\ll V_1$ where $V_1\ll V_0$ and that it smoothly matches
(\ref{tach}) at $X\sim V_1$. Then it is clear that the signal
velocity is exactly unity. The Lagrangian density (\ref{change})
describes a massless scalar field minimally coupled to gravity and a
cosmological constant.

The correspondingly corrected macroscopic gCg remains the same in
the observable region $X\gtrsim V_0$, or $\rho_{\rm Ch}+p_{\rm
Ch}\gtrsim V_0$ (note that $V_0$ is of the order of the present day
critical energy density in the Universe) but will become a mixture
of a cosmological constant and the extremely stiff ideal fluid
($p=\rho$) in the far future when $X$ will drop to $V_1$. With the
same accuracy, one can say that the corrected gCg becomes the usual
Chaplygin gas when its density is very close to $V_0$, much closer
than the one which occurs presently. Thus, by slightly changing the
gCg equation of state in the unobservable region, the causality
problem is cured. As a result, the present gCg superluminal sound
velocity is a {\em transient} phenomenon which disappears at
$t\to\infty$.

Finally, note that in the limit $\alpha\to\infty$, the gGg realizes
the Parker-Raval scenario \cite{PR99} of a sudden transition from
the matter--dominated stage with $a(t)\propto t^{2/3}$ to the de
Sitter expansion with a constant curvature (modulo a small term
$\propto \Omega_b$). The original idea of Parker and Raval to
produce such a scenario using non-perturbative vacuum polarization
of a light minimally coupled scalar field is questionable, both
regarding derivation of the effective action and stability of the
resulting specific kind of the $f(R)$ theory of gravity (here $R$ is
the Ricci scalar). But the same scenario based on the limiting case
of the gCg does not have these problems. This shows once more that
the superluminal gCg deserves further, more detailed study.

\ack The work of AYK and AAS was partially supported by the Research
Programme "Astronomy" of the Russian Academy of Sciences, by the
RFBR grant 05-02-17450 and by the grant LSS-1157.2006.2. AAS also
thanks Prof. Misao Sasaki and the Yukawa Institute for Theoretical
Physics, Kyoto University, for hospitality during the middle part of
this project.

\appendix

\section{Exact solution for perturbations of the gCg+baryons model at
the matter--dominated regime}

At the matter--dominated regime, the system (\ref{gensyssimplified})
describing gCg and baryons takes the following form:
\begin{equation}\label{SolStarsys}
\left\{
\begin{array}{l}
\ddot{\delta}_{\rm b} + \left(\frac{\dot{\h}}{\h} +
\frac{2}{a}\right)\dot{\delta}_{\rm b}
= \frac{1}{\h^{2}}\left(\rho_{b}\delta_{\rm b} + \rho_{Ch}\delta_{\rm Ch}
\right)\\
\ddot{\delta}_{\rm Ch} + \left(\frac{\dot{\h}}{\h} +
\frac{2}{a}\right)\dot{\delta}_{\rm Ch} +
\frac{k^{2}}{a^{2}\h^{2}}c_{\rm s}^{2}\dot{\delta}_{\rm Ch} =
\frac{1}{\h^{2}}\left(\rho_{b}\delta_{\rm b} + \rho_{Ch}\delta_{\rm
Ch}\right),
\end{array}
\right.
\end{equation}
where $c_{\rm s}^{2}$ is the gCg square sound velocity given by
(\ref{gcgw}) and (\ref{gcgcs}).

This system can be exactly solved using the technique developed
in \cite{SolovStar} for cosmological perturbations of two--fluid
models in the Newtonian approximation. Let us introduce the following
variables:
\begin{equation}
\begin{array}{ll}
x = k\gamma^{-2}a^{-\frac{3}{2}\gamma}, & \gamma = -2\alpha -
\frac{8}{3},
\end{array}
\end{equation}
where $\alpha$ is the gCg parameter. It is possible to extract from
system (\ref{SolStarsys}) a fourth order equation for $\delta_{\rm
Ch}$:
\begin{eqnarray}\label{SolStareq}
\left[\left(\Delta + \frac{2}{3\gamma}\right)\Delta\left(\Delta -
\frac{1}{3\gamma}\right)\left(\Delta - \frac{1}{\gamma}\right)
+\right.\nonumber\\
\left.x\left(\Delta^{2} + \frac{2\gamma - 1/3}{\gamma}\Delta +
\frac{1}{\gamma^{2}}\left(\gamma\left(\gamma - \frac{1}{3}\right) -
\frac{2}{3}\Omega_{\rm b0}\right)\right)\right]\delta_{\rm Ch} = 0,
\end{eqnarray}
where $\Delta = x\cdot{\rm d}/{\rm d}x$.

As found in \cite{SolovStar}, the general solution of
(\ref{SolStareq}) can be represented in terms of the Meijer
G--functions (the generalized hypergeometric functions):
\begin{eqnarray}\label{deltaChsolovstaro}
\delta_{\rm Ch} &=& C_{\rm 1}G_{\rm 24}^{\rm 41}\left(x|^{a_{\rm
1}a_{\rm 2}}_{b_{\rm 1}b_{\rm 2}b_{\rm 3}b_{\rm 4}}\right) + C_{\rm
2}G_{\rm 24}^{\rm 41}\left(x|^{a_{\rm 2}a_{\rm 1}}_{b_{\rm 1}b_{\rm
2}b_{\rm 3}b_{\rm 4}}\right) + \nonumber\\ &+& C_{\rm 3}G_{\rm
24}^{\rm 40}\left(x\rme^{\rmi\pi}|^{a_{\rm 1}a_{\rm 2}}_{b_{\rm
1}b_{\rm 2}b_{\rm 3}b_{\rm 4}}\right) + C_{\rm 4}G_{\rm 24}^{\rm
40}\left(x\rme^{-\rmi\pi}|^{a_{\rm 1}a_{\rm 2}}_{b_{\rm 1}b_{\rm
2}b_{\rm 3}b_{\rm 4}}\right),
\end{eqnarray}
where:
\begin{equation}
\begin{array}{lll}
a_{\rm 1,2} = \frac{1}{6\gamma}\left(1\mp\sqrt{1 +
24\Omega_{\rm b0}}\right)\\
b_{\rm 1} = -\frac{2}{3\gamma}, & b_{\rm 2} = 0\\ b_{\rm 3} =
\frac{1}{3\gamma}, & b_{\rm 4} = \frac{1}{\gamma},
\end{array}
\end{equation}
the $C_{i}$ ($i=1,2,3,4$) are integration constants and the Meijer
G--functions are defined as follows \cite{Erde}:
\begin{equation}\label{MeijerG}
\fl G_{p,q}^{m,n}\left(x|^{a_{\rm
1},...,a_{p}}_{b_{1},...,b_{q}}\right) = \frac{1}{2\pi
\rmi}\int_{\gamma_{L}}\frac{\prod_{j=1}^{m}\Gamma(b_{j} - s)
\prod_{j=1}^{n}\Gamma(1 - a_{j} + s)}{\prod_{j=n+1}^{p}\Gamma(a_{j}
- s) \prod_{j=m+1}^{q}\Gamma(1 - b_{j} + s)}x^{s}ds,
\end{equation}
where $\Gamma(s)$ is the Gamma function and the integration contour
$\gamma_{\rm L}$ circuits all the poles of $\Gamma(1 - a_{j} + s)$
anti--clockwise and all the poles of $\Gamma(b_{j} - s)$ clockwise
(all these poles lie along the real axis).

The solution for $\delta_{\rm b}$ has the same functional form as
(\ref{deltaChsolovstaro}), with:
\begin{equation}
a_{\rm 1,2} = 1 + \frac{1}{6\gamma}\left(1\mp\sqrt{1 + 24\Omega_{\rm
Ch0}}\right),
\end{equation}
where $\Omega_{\rm Ch0} = 1 - \Omega_{\rm b0}$ since we assume
a spatially flat universe.

\end{document}